\documentclass[a4paper]{article}

\usepackage{INTERSPEECH2021}
\usepackage{csquotes}
\usepackage{gensymb}
\usepackage{amsmath}
\usepackage{amssymb}
\usepackage{xcolor}
\usepackage{booktabs}
\usepackage{multirow}
\usepackage{booktabs}
\usepackage{multirow}
\usepackage{url}
\urldef{\ultraxurl}\url{http://www.ultrax-speech.org}

\usepackage{hyperref}

\title{Silent versus modal multi-speaker speech recognition \\ from ultrasound and video}

\name{Manuel Sam Ribeiro$^{1}$\thanks{\hspace*{-.5em}$^{1}$Manuel Sam Ribeiro is now at Amazon.}, 
Aciel Eshky$^{2}$\thanks{\hspace*{-.5em}$^{2}$Aciel Eshky is now at Rasa Technologies.},
Korin Richmond,
Steve Renals}
\address{The Centre for Speech Technology Research, University of Edinburgh, UK}
\email{\{sam.ribeiro, aeshky, korin.richmond, s.renals\}@ed.ac.uk}

\begin{document}

\maketitle

\begin{abstract}
We investigate multi-speaker speech recognition from ultrasound images of the tongue and video images of the lips.
We train our systems on imaging data from modal speech, and evaluate on matched test sets of two speaking modes: silent and modal speech.
We observe that silent speech recognition from imaging data underperforms compared to modal speech recognition, likely due to a speaking-mode mismatch between training and testing.
We improve silent speech recognition performance using techniques that address the domain mismatch, such as fMLLR and unsupervised model adaptation.
We also analyse the properties of silent and modal speech in terms of utterance duration and the size of the articulatory space.
To estimate the articulatory space, we compute the convex hull of tongue splines, extracted from ultrasound tongue images.
Overall, we observe that the duration of silent speech is longer than that of modal speech, and that silent speech covers a smaller articulatory space than modal speech.
Although these two properties are statistically significant across speaking modes, they do not directly correlate with word error rates from speech recognition.
\end{abstract}

\section{Introduction}

Silent speech is a speaking mode characterised by the absence of an audible speech signal \cite{schultz2017biosignal}.
When silently articulating, speakers activate their oral and nasal articulators, but suppress laryngeal activity, thus producing no audible output.
The speech production process must therefore be captured by directly measuring the activity of the vocal tract articulators (e.g. lips, tongue).
Silent speech interfaces (SSIs) perform speech recognition and synthesis from such articulatory measurements in order to restore spoken communication for users with voice impairments or to allow silent communication in situations where audible speech is undesirable \cite{denby2010silent}.

Articulatory movement can be captured \cite{schultz2017biosignal} using \emph{articulography techniques}, such as electromagnetic articulography (EMA) or permanent-magnetic articulography (PMA); \emph{palatography techniques}, such as electropalatography or optopalatography; or \emph{imaging techniques}, such as video imaging, magnetic resonance imaging (MRI) or ultrasound tongue imaging (UTI).
Video is the most convenient method to acquire articulatory imaging data.
Its main limitation is its inability to capture anything beyond extraoral articulators, such as the lips and jaw.
To provide complementary data, intraoral articulators can be monitored using medical imaging techniques such as MRI \cite{scott2014speech} and UTI \cite{stone2005guide}. 
Although MRI provides high-quality images, it is expensive, not easily accessible, and suffers from loud background noise, a supine recording position, and low temporal resolution. 
Ultrasound, on the other hand, is relatively cheap, portable, non-invasive, non-intrusive, and clinically safe \cite{stone2005guide}. UTI uses diagnostic ultrasound operating in B-mode to visualise the tongue surface during speech production. A real-time B-mode ultrasound transducer is placed under the speaker’s chin to generate a midsaggital or coronal view of the tongue. 
There are, however, some challenges associated with ultrasound images \cite{stone2005guide, ribeiro2019speaker}.
UTI output tends to be noisy, with unrelated high-contrast edges, speckle noise, or interruptions of the tongue surface.
Image quality tends also to be affected by speaker characteristics (e.g. age and physiology) and session variability (e.g. incorrect or variable probe placement).

Silent speech interfaces using ultrasound and video imaging for silent speech recognition have been proposed in previous studies \cite{hueber2010development, florescu2010silent, tatulli2017feature, ji2018updating, kimura2019sottovoce, kimura2020end}.
Standard speech recognition techniques are typically applied to silent speech, either introducing methods to extract and combine features from visual modalities \cite{tatulli2017feature, liu2016comparison} or investigating acoustic model architectures \cite{hueber2010development, ji2018updating, kimura2019sottovoce, hueber2009visuo}.
Most studies have used small single-speaker databases, such as the Silent Speech Challenge dataset \cite{cai2011recognition}, which contains roughly 2500 utterances of ultrasound and video images from a single native English speaker.
Recent work, however, has shown that multi-speaker and speaker-independent ultrasound-based SSIs are a challenging problem \cite{ribeiro2019speaker, ribeiro2021tal}.

Although we are primarily concerned with silent speech, there are other speaking modes \cite{schultz2017biosignal}.
\emph{Modal speech}, alternatively termed \enquote{vocalised} or \enquote{phonated} speech, denotes a standard speaking mode where laryngeal and pulmonary activity operate as normal.
Related modes include \emph{shouted}, \emph{whispered}, or \emph{murmured speech}, which allow laryngeal activity with varying airflow intensity.
\emph{Imagined} speech and \emph{inner} speech denote speaking modes that do not require articulatory movement, and therefore must rely upon biosignals captured at the neural level with Brain-Computer Interfaces  \cite{wolpaw2002brain, herff2016automatic, clayton2020decoding}.

Previous work has investigated and compared the articulatory properties of silent, whispered, and modal speaking modes.
Studies have measured tongue and lip movements with EMA \cite{janke2010impact, dromey2017effects, teplansky2019tongue, teplansky2020tongue} or with ultrasound and video imaging \cite{crevier2011articulatory}.
Two conflicting hypotheses have arisen with regard to silent articulation \cite{dromey2017effects}.
The first claims that the absence of laryngeal activity and auditory feedback forces speakers to rely on somato-sensory feedback to find the correct placement for their articulators, thus leading to hyper-articulation.
The second claims that the lack of intra-oral pressure (e.g.\ for plosives or fricatives) causes speakers to produce incomplete or reduced movements, thus leading to hypo-articulation.
Recent work has found evidence to support the latter hypothesis \cite{dromey2017effects, teplansky2019tongue, teplansky2020tongue, crevier2011articulatory}.
These studies have observed that silent speech exhibits longer duration and decreased articulatory activity when compared with modal speech.
Similarly, whispered speech has been found to have reduced duration \cite{dromey2017effects, teplansky2019tongue} when compared with modal speech.
The articulatory space of whispered speech was found to be significantly smaller than that of modal speech, but larger than that of silent speech \cite{teplansky2019tongue}.
However, evidence for hyper-articulation was also observed \cite{janke2010impact}, which might occur only for specific speech sounds such as bilabial consonants.
These studies relied on a relatively small number of participants, typically 4-12 speakers \cite{florescu2010silent, teplansky2019tongue, teplansky2020tongue, janke2010impact}, with one study using 20 speakers \cite{dromey2017effects}.

This paper investigates a \emph{multi-speaker ultrasound-based silent speech interface and the articulatory properties of modal and silent speech across 82 speakers}.
Our key contributions are:
1) an investigation of multi-speaker and speaker-dependent modal and silent speech recognition;
2) a description of the articulatory properties of modal and silent speech in terms of duration and articulatory space; and
3) an analysis of speech recognition results in terms of the observed articulatory properties of modal and silent speech.

\section{The TaL corpus}
\label{sec:data}
The Tongue and Lips corpus \cite{ribeiro2021tal} is a multi-speaker corpus of audio, ultrasound tongue imaging, and lip videos.
The corpus is distributed with the Ultrasuite Repository\footnote{\label{fn:ultrasuite}\url{https://www.ultrax-speech.org/ultrasuite}} \cite{eshky2018ultrasuite}.
TaL contains two subsets:
\textbf{TaL1} has six recording sessions from a professional voice talent and male native English speaker; and
\textbf{TaL80} has single session recordings from 81 native English speakers without voice talent experience.
In total, the TaL corpus has approximately 24 hours of synchronised parallel ultrasound, video, and audio, of which  13.5 hours are speech.
Ultrasound in the TaL corpus was recorded using Articulate Instruments' Micro system \cite{articulate2010articulate} at $\sim$80 fps with a 92\degree ~field of view.
Each \enquote{raw} ultrasound frame contains 842 echo returns for each of 64 scan lines ($64\times842$) and captures a midsaggital view of the tongue.
Lip videos were recorded at $\sim$60 fps in grayscale and resized to $240\times320$ pixels.
Audio was captured with a Sennheiser HKH 800 p48 microphone with a 48KHz sampling frequency at bit depth of 16-bit.

\section{Experiments}
\label{sec:experiments}

\subsection{Data preparation}

The TaL corpus includes modal and silent speech utterances, with TaL1 also including whispered speech.
Using the prompt text, we define matching \textbf{test sets} across speaking modes.
That is, test data contains the same set of sentences read silently and audibly.
For TaL1, we include an additional corresponding test set of whispered speech.
There are 1374 utterances for each TaL80 test set.
This includes 2.72 and 2.30 hours of data for the silent and modal speech test sets, respectively.
For the TaL1 test data, there are 106 utterances for each test set, corresponding to 10-11 minutes of data.

The TaL corpus contains primarily modal speech, therefore we define our \textbf{training set} over those utterances, excluding all prompts already occurring in the test data.
We consider two training scenarios.
The \textbf{multi-speaker} scenario uses data from all 82 speakers.
TaL1 contains more data than each speaker in TaL80, so we randomly select 145 utterances across recording sessions.
The multi-speaker training set contains approximately 11700 utterances, with an average of 136 utterances per speaker.
In total, there are 16.4 hours for training, with an average of 12 minutes of data for each of the 82 speakers.
The \textbf{speaker-dependent} scenario uses data from TaL1.
We discard \emph{day1}, for which video synchronisation is not available \cite{ribeiro2021tal}.
This training set contains 935 utterances across 5 recording sessions, with an average of 187 utterances per session.
In total, there are 1.42 hours for training, with an average of 17 minutes per session.

\subsection{Feature extraction}

\begin{figure}[t]
  \centering
  \includegraphics[width=.92\linewidth]{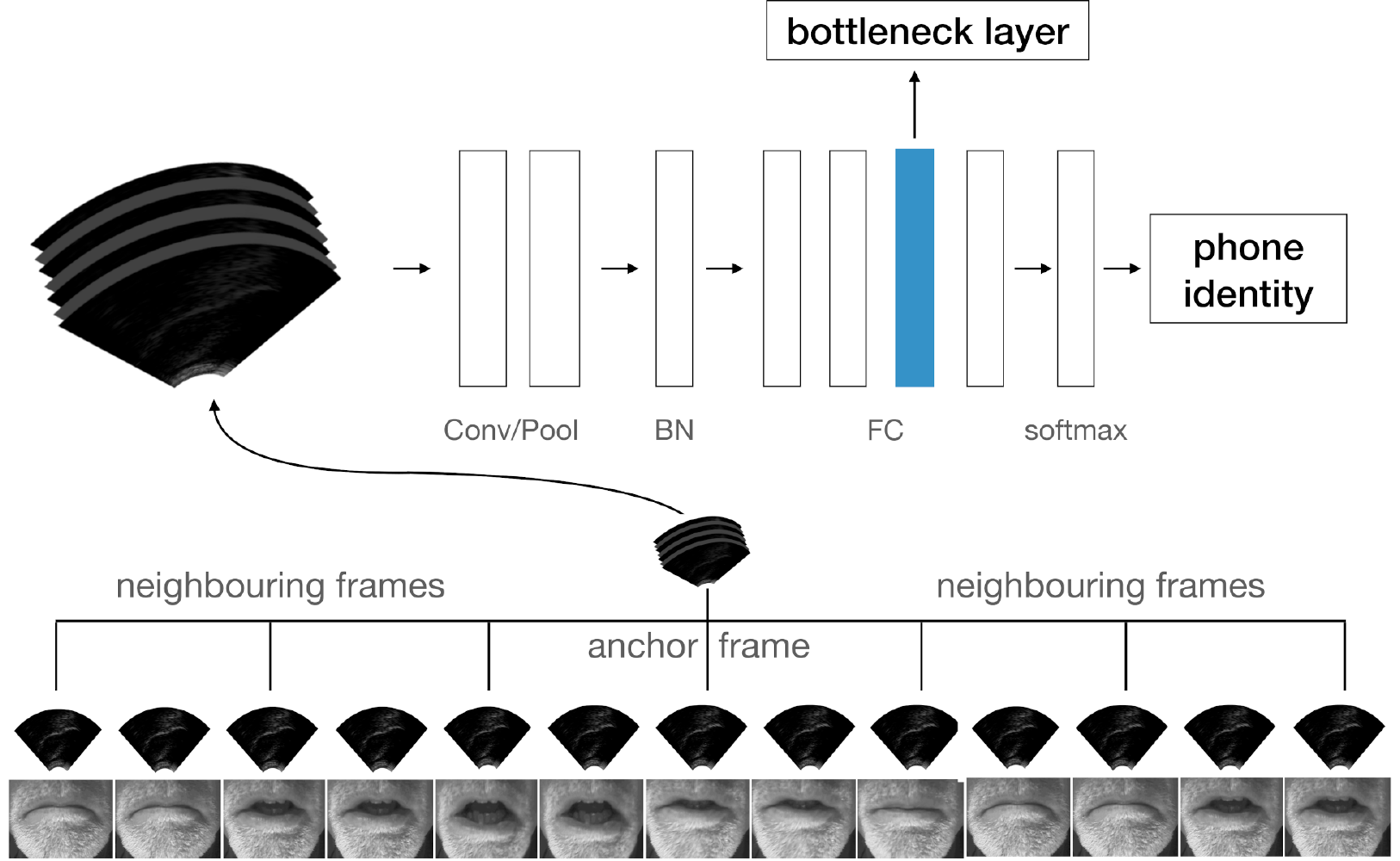}
  \caption{Feature extraction network.}
  \label{fig:feature-extraction-diagram}
\end{figure}

The training data consists of modal speech, so we exploit the audio stream to bootstrap a simple supervised feature extractor.
The Kaldi speech recognition toolkit \cite{povey2011kaldi} is used to force-align the training data at the phone level using audio.
The time boundaries are then matched to the visual streams, which allows our feature extraction model to be optimised on phone identities.

Using bi-linear interpolation, ultrasound and video frames are resized to $64 \times 128$ and $120 \times 160$, respectively.
Video frames are then further cropped to $64 \times 128$.
The data is normalised to zero mean and unit variance.
A single input sample consists of an anchor frame and a set of neighbouring frames.
Neighbouring frames are extracted over windows of 12 frames with a frame shift of 4 frames.
Each sample therefore consists of 7 frames, grouped as multiple channels: the anchor frame and 3 left and right neighbouring frames.
Input frame sequences are processed by two convolutional layers, using 10x10 kernels with 64 and 128 filters, respectively, and ReLu activation functions.
Each layer is followed by max-pooling with a 2x2 kernel.
One-dimensional batch normalisation is then applied on the flattened output of the second convolutional layer.
The normalised feature vector is further processed by four fully-connected layers with rectified linear units. 
Each layer contains 1024, 512, 128, 512 nodes, respectively.
A final fully-connected layer is followed by the softmax objective function.
The number of output classes (number of unique phones) is 49.
Models are optimized via Stochastic Gradient Descent with minibatches of 256 samples and an L2 regulariser with weight 0.1.
Training runs for 30 epochs with a learning rate of 0.001.
After each epoch, the model is evaluated on the validation data and we keep the best model across all epochs.
Figure \ref{fig:feature-extraction-diagram} illustrates the feature extraction network.

We call the third fully-connected layer with 128 hidden units the \emph{bottleneck layer}.
Ultrasound and video samples are processed with the trained networks to extract bottleneck features.
The output is a 128 dimensional feature vector for each frame in the TaL corpus.
We have made no attempt at this stage to optimize our feature extraction method, as this is not the primary goal of the paper.
Earlier work has shown that complex feature extraction approaches can lead to substantial improvements \cite{tatulli2017feature, liu2016comparison}.
Our method remains fairly straightforward and we leave further optimizations for future work.

\subsection{Systems and results}

\textbf{Systems} are trained using Kaldi \cite{povey2011kaldi}, with models being initialised from a flat start using the bottleneck features.
After monophone and triphone training, input features are processed with Linear Discriminant Analysis (LDA) and a Maximum Likelihood Linear Transform  (MLLT).
This is followed by Speaker Adaptive Training (SAT) with feature-space MLLR (fMLLR \cite{rath2013improved}).
In the speaker-dependent scenario, each recording session is treated as a separate speaker for SAT.
The alignment and HMM models from this stage are then used to train a time-delay neural network (TDNN, \cite{peddinti2015time}) following Kaldi's \emph{nnet3} recipe with a learning rate of $0.0015$.
We use \enquote{raw} bottleneck features or \enquote{fMLLR} features as input to the TDNN.
Decoding is done with a simple in-domain bigram language model trained on a superset of the prompts used to record the TaL corpus \cite{ribeiro2021tal}.
We further investigate an unsupervised adaptation scenario, where the decoding output of a baseline model is used to guide fine-tuning
We adapt all parameters of the model on targets from each test set separately.

\begin{table}[t]
\centering
\caption{Word error rate on modal, silent, and whispered speech test sets for TaL1 and TaL80. Results are shown for systems using either \enquote{raw} bottleneck features or fMLLR features with a TDNN acoustic model. Results for unsupervised model adaptation on each test set include in parenthesis the difference in WER over the corresponding baseline model.}
\label{tab:wer-results}
\resizebox{1.0\columnwidth}{!}{%
\begin{tabular}{@{}lcc|cc@{}}
\toprule
\multicolumn{1}{c}{\textbf{Test Set}} & \textbf{Raw}     & \textbf{fMLLR}                       & \textbf{Raw}                 & \textbf{fMLLR}              \\ \midrule
\multicolumn{3}{c}{\textbf{multi-speaker}}     & \multicolumn{2}{c}{\textbf{+ unsupervised adapt}} \\  \midrule[.02em]
\multicolumn{5}{c}{\textit{TaL80}}                                                                                   \\  \midrule[.02em]
\textbf{modal}                    & 39.34   & 39.79                       & 41.76 \enspace \textit{(+2.42)}       & 41.37 \enspace \textit{(+1.58)}      \\
\textbf{silent}                   & 77.79   & 70.21                       & 75.83 \enspace \textbf{\textit{(-1.96)}}       & 69.84 \enspace \textbf{\textit{(-0.37) }}     \\  \midrule[.02em]
\multicolumn{5}{c}{\textit{TaL1}}                                                                                    \\  \midrule[.02em]
\textbf{modal}                    & 44.76   & 46.21                       & 44.86 \enspace \textit{(+0.10)}       & 47.84 \enspace \textit{(+1.63)}      \\
\textbf{silent}                   & 56.87   & 50.05                       & 54.18 \enspace \textbf{\textit{(-2.69)}}       & 48.32 \enspace \textbf{\textit{(-1.73)} }     \\
\textbf{whispered}                & 42.36   & 38.71                       & 42.76 \enspace \textit{(+0.40) }      & 40.54 \enspace \textit{(+1.83)}      \\ \midrule
\multicolumn{3}{c}{\textbf{speaker-dependent}} & \multicolumn{2}{c}{\textbf{+ unsupervised adapt}} \\ \midrule[.02em]
\multicolumn{5}{c}{\textit{TaL1}}                                                                                    \\ \midrule[.02em]
\textbf{modal}                    & 33.33   & 29.59                       & 34.39 \enspace \textit{(+1.06)}       & 30.84 \enspace \textit{(+1.25)}     \\
\textbf{silent}                   & 52.64   & 38.42                       & 49.28 \enspace \textbf{\textit{(-3.36)}}       & 37.94 \enspace \textbf{\textit{(-0.48)}}      \\
\textbf{whispered}                & 29.68   & 25.07                       & 31.12 \enspace \textit{(+1.44)}       & 25.74 \enspace \textit{(+0.67)}      \\ 
\bottomrule
\end{tabular}%
}
\end{table}

Table \ref{tab:wer-results} shows word error rate (WER) \textbf{results} for all systems.
We observe that in all scenarios performance on silent speech is substantially lower than on modal speech.
This is unsurprising, as training is done on modal speech data.
Previous experiments have shown similar results, where a mismatch of speaking mode between training and testing time led to poor results \cite{florescu2010silent}.
A degradation when training on modal speech and testing on silent speech was also observed on a speech reconstruction task \cite{zhang2020tal}.
Our results further demonstrate that strategies which model the mismatch between training and test domains, such as fMLLR or unsupervised model adaptation, improve the performance of the silent speech test set.
These techniques have a negative impact on the modal speech test sets, as this type of data was already seen at training time.
For TaL1, results on the speaker-dependent system are better than those of the multi-speaker system, even though there is less data overall.
This further shows the challenges of processing data from multiple speakers.
We observe a positive impact of fMLLR in the speaker-dependent results.
This is likely due to session variability, as we learn a separate fMLLR transform for each recording session.
Whispered speech in the TaL1 sets performs better than modal speech.
This is surprising, as previous work showed that mismatched conditions can also lead to poor recognition performance on whispered speech \cite{srinivasan2019study}, although using EMA.
It might also be that these results are specific to the TaL1 speaker.
In the following section, we investigate the properties of modal, whispered, and silent speech data in the TaL corpus.

\section{Analysis}
\label{sec:analysis}

\begin{figure}[t]
  \centering
  \includegraphics[width=\linewidth]{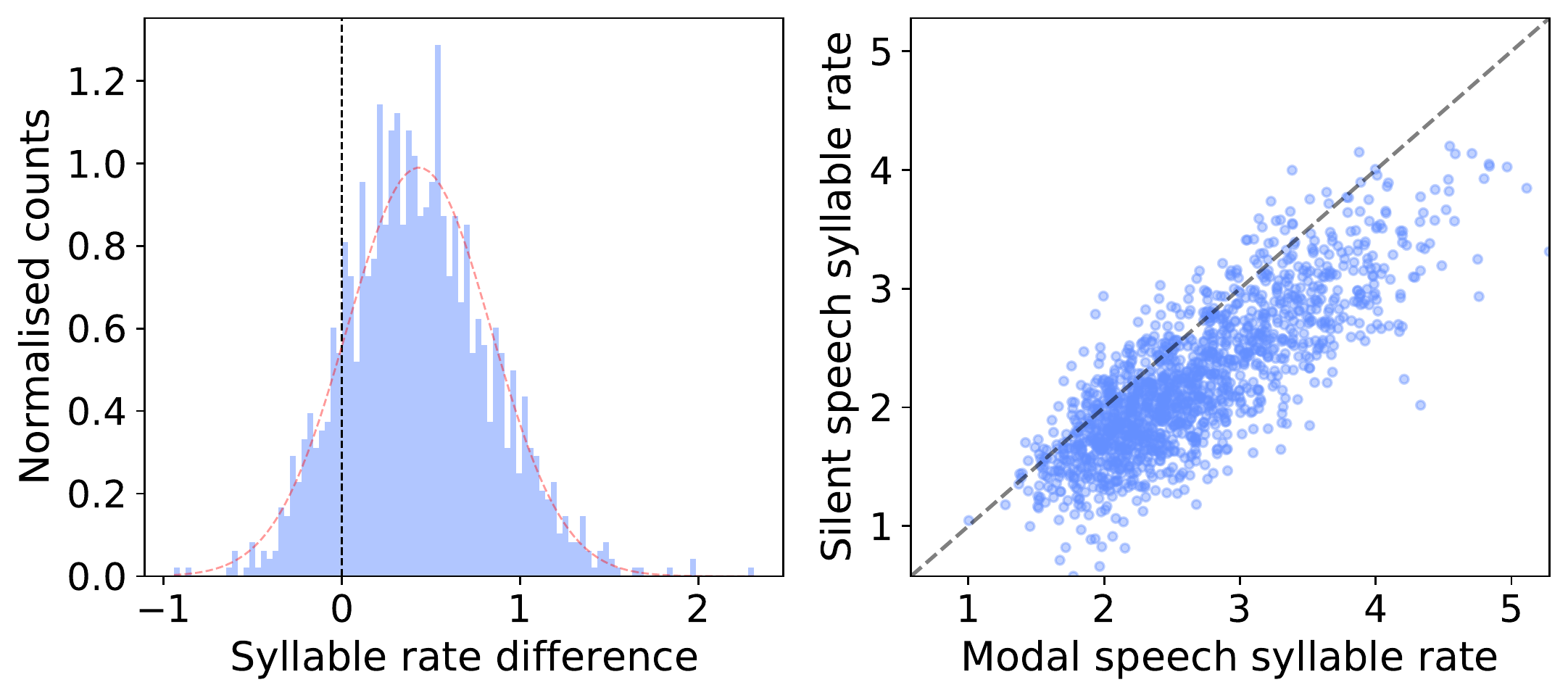}
  \caption{Syllable rate for modal and silent speech utterances in the TaL80 test sets. Figure on the left shows histogram of syllable rate difference across matched utterances. Figure on the right shows the paired utterance syllable rates.}
  \label{fig:syl-rate-hist-scatter}
\end{figure}

For the analysis of speaking modes, we consider utterance duration, measured in terms of syllable rate; and articulatory space, measured in terms the area of the convex hull given by tongue splines extracted from ultrasound images of the tongue.

In terms of \textbf{utterance duration}, we conduct a two-tailed paired t-test on the syllable rate for each utterance in the TaL80 modal and silent speech test sets.
We observe a significant difference ($p < .001$) in the syllable rate for modal ($\mu=2.66, \sigma=0.68$) and silent ($\mu=2.21, \sigma=0.61$) speech conditions.
Figure \ref{fig:syl-rate-hist-scatter} shows the distribution of syllable rate for the utterances in the TaL80 test sets.
A similar test is conducted for each utterance in the TaL1 modal, whispered, and silent test sets.
To account for multiple comparisons, we perform a Holm-Bonferroni correction on all results.
We compute the syllable rate for modal ($\mu=2.31, \sigma=0.55$), whispered ($\mu=2.20, \sigma=0.54$), and silent ($\mu=2.11, \sigma=0.53$) speech conditions.
We observe that the syllable rate for the three speaking modes are significantly different at the level of $p < .001$.
These results support those found in previous studies \cite{dromey2017effects, teplansky2019tongue, teplansky2020tongue, crevier2011articulatory}, indicating that silent speech is articulated slower than whispered speech, and that whispered speech is slower than modal speech.

\begin{figure}[t]
  \centering
  \includegraphics[width=\linewidth]{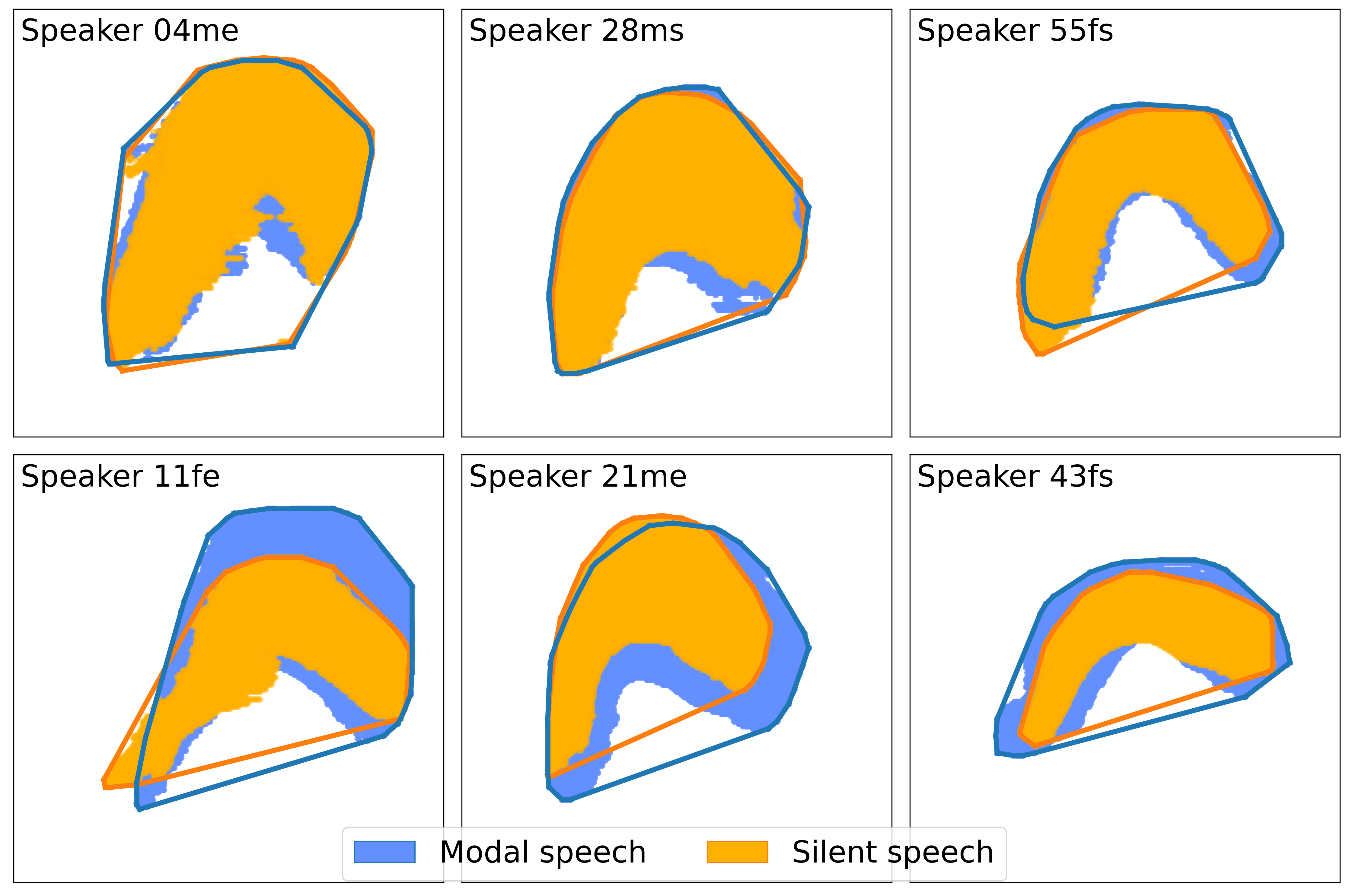}
  \caption{Articulatory space for modal and silent speech. Tongue contours are identified by MTracker \cite{zhu2018automatic} and pruned with Isolation Forests \cite{liu2008isolation}. The darker line indicates the convex hull of the set of tongue contours. The top row shows speakers with a small difference in convex hull area, while the bottom row shows speakers with a larger difference.}
  \label{fig:hull-area-speaker-samples}
\end{figure}

\begin{figure}[t]
  \centering
  \includegraphics[width=\linewidth]{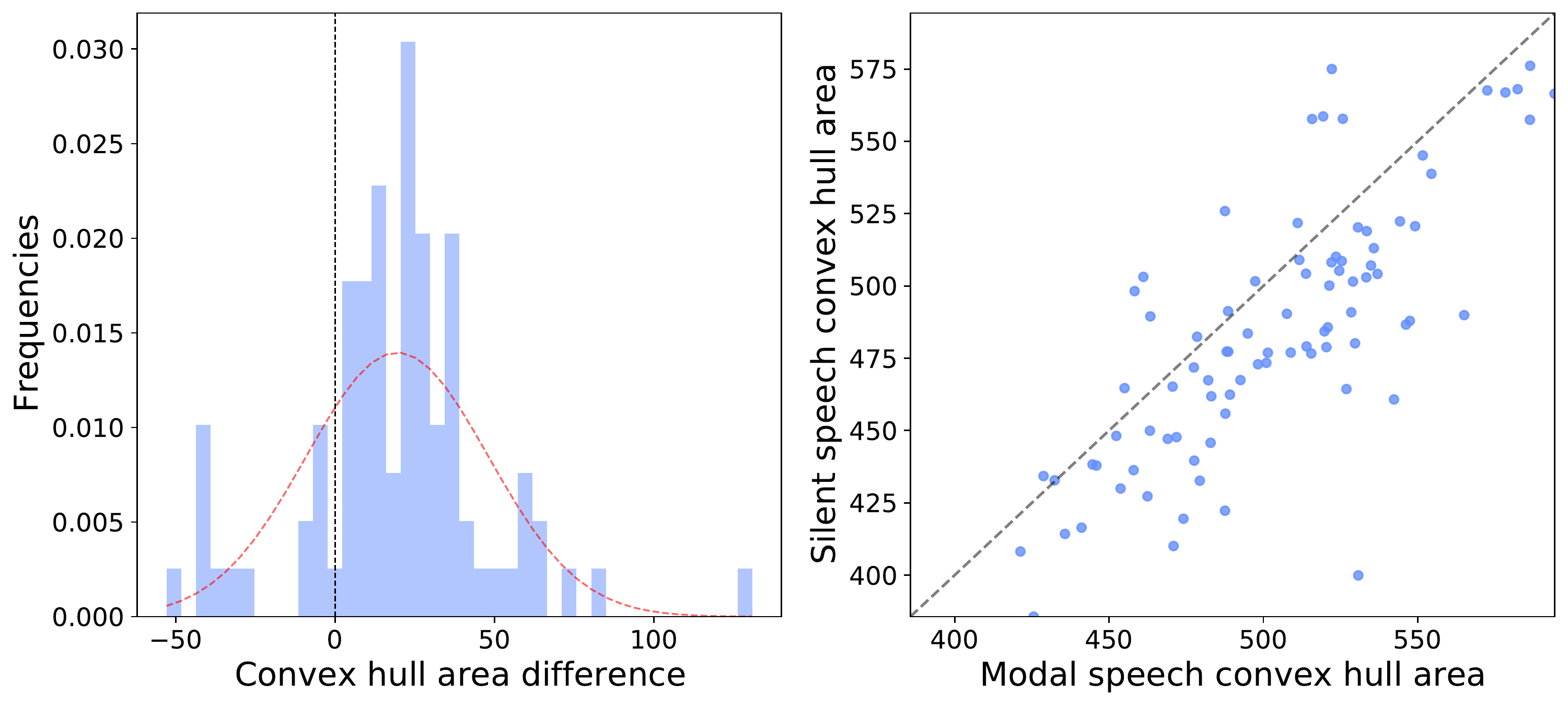}
  \caption{Convex hull area for modal and silent speech speakers in the TaL80 test sets. Figure on the left shows histogram of convex hull area difference. Figure on the right shows the paired speaker convex hull areas.}
  \label{fig:hull-area-hist-scatter}
\end{figure}

To measure \textbf{articulatory space}, we use MTracker \cite{zhu2018automatic} to extract tongue contours for the ultrasound images.
The output of MTracker is further post-processed using Isolation Forests \cite{liu2008isolation}, estimated separately for the set of utterances belonging to each speaker in each test set.
Isolation Forests are an unsupervised approach frequently used for anomaly detection.
This post-processing approach removes outliers from the \enquote{tongue pixels} identified by the tracker, which often correspond to noise in the ultrasound images.
We then compute the convex hull and its area for each speaker and speaking mode separately.
Figure \ref{fig:hull-area-speaker-samples} illustrates the set of tongue contours for six speakers in TaL80.
Two-tailed paired t-tests are used to analyse the difference in convex hull areas between speaking modes.
We find a significant difference ($p < .001$) in the convex hull area for modal ($\mu=501.64, \sigma=39.89$) and silent ($\mu=480.93, \sigma=42.86$) speech conditions.
Figure \ref{fig:hull-area-hist-scatter} shows the distribution of convex hull areas for the speakers in the TaL80 test sets.
We repeat the process for TaL1, considering the convex hull area for modal ($\mu=537.28, \sigma=24.31$), whispered ($\mu=537.29, \sigma=28.10$), and silent ($\mu=536.99, \sigma=28.67$) speech conditions. We observe that the convex hull area is not significantly different across the three speaking modes for the TaL1 speaker.
Overall results suggest that speakers, on average, tend to hypo-articulate when silently articulating.
However, we do find a small number of speakers that do not exhibit such behaviour.
For TaL1, this may be due to the speaker's voice talent experience.
Although most speakers hypo-articulate on average, further work can investigate whether this behaviour is consistent for all phones.
Previous work found evidence of hyper-articulation for specific phone classes \cite{janke2010impact}.
Future work may also consider lip movement.

\begin{figure}[t]
  \centering
  \includegraphics[width=\linewidth]{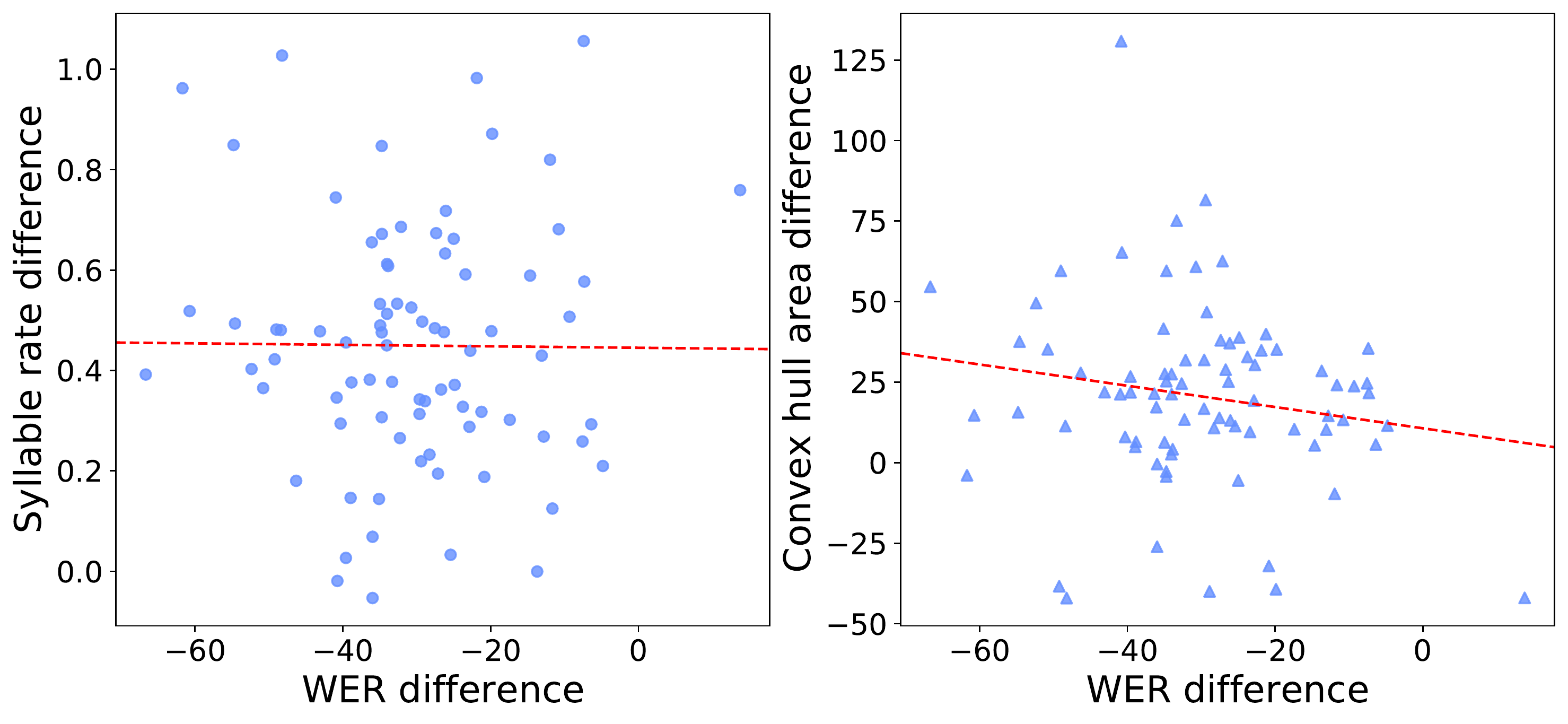}
  \caption{Difference between modal and silent speech in terms of word error rate, syllable rate, and convex hull area. Each sample represents one speaker in the TaL80 dataset. The dashed red line indicates a best linear fit to the samples.}
  \label{fig:wer-sylrate-hullarea-scatter}
\end{figure}

Finally, we compare utterance duration and size of articulatory space with the results obtained from the speech recognition systems.
We compute \textbf{word error rate} separately for each speaker in the TaL80 dataset using the multi-speaker system with fMLLR, which has the best results on average.
To compare the two speaking modes, we take the difference between measurements for modal and silent speech.
Figure \ref{fig:wer-sylrate-hullarea-scatter} illustrates the relationship between syllable rate and convex hull area with respect to word error rate.
We find no correlation between syllable rate difference and WER differences ($r=-.009$).
And we find a weak negative correlation between convex hull area difference and WER difference ($r=-.1682$).
Similarly, we find a weak negative correlation between syllable rate difference and convex hull area difference ($-.132$).
This is a surprising finding:
\emph{Although there are significant differences in utterance duration and articulatory space between modal and silent speech, they do not help explain speech recognition performance between the two speaking modes}.
The observed difference in WER scores might partly be explained due to speaker and session differences.
For example, it was previously observed that some speakers consistently perform well across tasks, while others do not \cite{ribeiro2021tal}.
Additionally, it is possible that the speakers do not articulate correctly when deprived of audio feedback.
Further work might attempt to quantify the error in the silent speech utterances.
Rather than relying on domain adaptation, additional collection of silent speech for training might help reduce the gap between modal and silent speech recognition results.
Those findings should be useful when developing speaker-independent silent speech recognition.

\section{Conclusion}

We investigated a multi-speaker silent speech interface from ultrasound images of the tongue and video images of the lips.
The systems were trained on modal speech and evaluated on matched test sets of silent and modal speech.
Performance on silent speech was substantially lower than on modal speech, indicating a domain mismatch between speaking modes.
Techniques that model that mismatch, such as fMLLR or unsupervised model adaptation, improve the results on silent speech but not on modal speech.
By analysing the characteristics of the speaking modes, we observed that silent articulation exhibits longer duration when compared to modal speech.
Similarly, the overall articulatory space in terms of tongue movement is in general smaller in silent speech than in modal speech, although some speakers in contrast hyper-articulate.
Although there are significant differences in terms of duration and articulatory space, they do not directly correlate with WER.

\medskip
\textbf{Acknowledgements}
Supported by the Carnegie Trust for the Universities of Scotland (Research Incentive Grant number 008585)
and the EPSRC Healthcare Partnerships grant number EP/P02338X/1 (Ultrax2020 –- \ultraxurl).

\bibliographystyle{IEEEtran}
\bibliography{references}

\end{document}